\documentclass[10pt]{article}
\usepackage{courier}
\usepackage{graphicx}
\usepackage{hyperref}
\usepackage{xcolor}
\hypersetup{
    colorlinks,
    linkcolor={red!50!black},
    citecolor={blue!50!black},
    urlcolor={blue!80!black}
}

\title{A Typology of Authentication Systems}
\author{Christopher S. Pilson \\
	Braintrust Applied Research, LLC \\
	\and 
	James C. McElroy \\
	Iowa State University \\
	}

\date{September 2, 2015}
	
\begin{document}

\maketitle

\section{Introduction}
Authentication systems are designed to give the right person access to an organization's information system and to restrict it from the wrong person. Such systems are designed by IT professionals to protect an organization's assets (e.g., the organization's network, database, or other information). Too often, such systems are designed around technical specifications without regard for the end user. We argue that doing so may actually compromise a system's security. The purpose of this paper, then, is to examine authentication systems from both the point of view of the organization and that of the user.

Research has examined and subsequently framed security flaws in a taxonomy \cite{Landwehr1994, Weber2005}, but the discussion surrounding security mechanisms has been technical in nature as in Anderson's formula \cite{Anderson1972} to determine password strength or the Receiver Operating Characteristics (ROC) curve. The ROC curve demonstrates the performance of an authentication system through plotting false acceptance rates (allowing an intruder access to the system) against false rejection rates (failing to correctly authenticate a valid system user) \cite{FidelicaMicrosystems2001}.

ROC curves, as a traditional description of an authentication mechanism's security, provide insight into how precise a particular authentication mechanism is, yet they provide no information about this mechanism as experienced by a user. In this paper, we present a typology of authentication systems that accounts for both system and user requirements. We then examine and classify existing authentication mechanisms in light of the typology, and end the paper with a call for greater consideration of the human element in human-computer access applications.

\section{Authentication System Types}
An authentication system must be able to differentiate an authentic user from an attacker. This aspect of authentication systems focuses on the overall security provided to a protected asset and can range from open access, such as a public library where anyone can access the asset, to highly secure military information systems. Consequently, the level of security associated with authentication systems traditionally varies in direct relation to the protected asset, ranging from very low (i.e., trusted computing), to very high (e.g., a system requiring a retinal scan as verification of user identity). The specific authentication mechanism used by an organization involves a trade-off between initial and ongoing authentication costs and the value of the protected asset. Authentication system security, therefore, forms the first dimension of our typology, as this is a necessary element of any system installed to protect an asset. However, because humans interact with computers, security alone seldom yields a full accounting of the utility of authentication systems.

Ease of use constitutes a second area of concern for any authentication system. Authentication mechanisms that are easy to use require little cognitive effort from a valid user. Trusted computing mechanisms, for example, involve granting system access using network or physical location. That is, anyone at a given location has authorization to directly log into the system. Other easy-to-use systems might require a simple password. Conversely, some systems require passwords that are either elaborate and/or mutable, or even phrases in order to obtain access to the system. Such systems, however, require users to remember these complex or mutable passwords. This places a high cognitive load upon the user. Thus, cognitive complexity of the authentication system constitutes the second dimension of our typology. This dimension is important because if the authentication system is too cognitively complex, users will develop shortcuts to alleviate these cognitive demands, such as writing their password on a Post-It\textsuperscript{TM} and attaching it to their computer screen. The result of such short-cuts is that the technically secure authentication system developed by the IT professional is compromised by the behavior of the user.

\begin{figure}[h!]
  \caption{Proposed Typology of Authentication Systems, with Exemplars}
  \centering
    \includegraphics[scale=1]{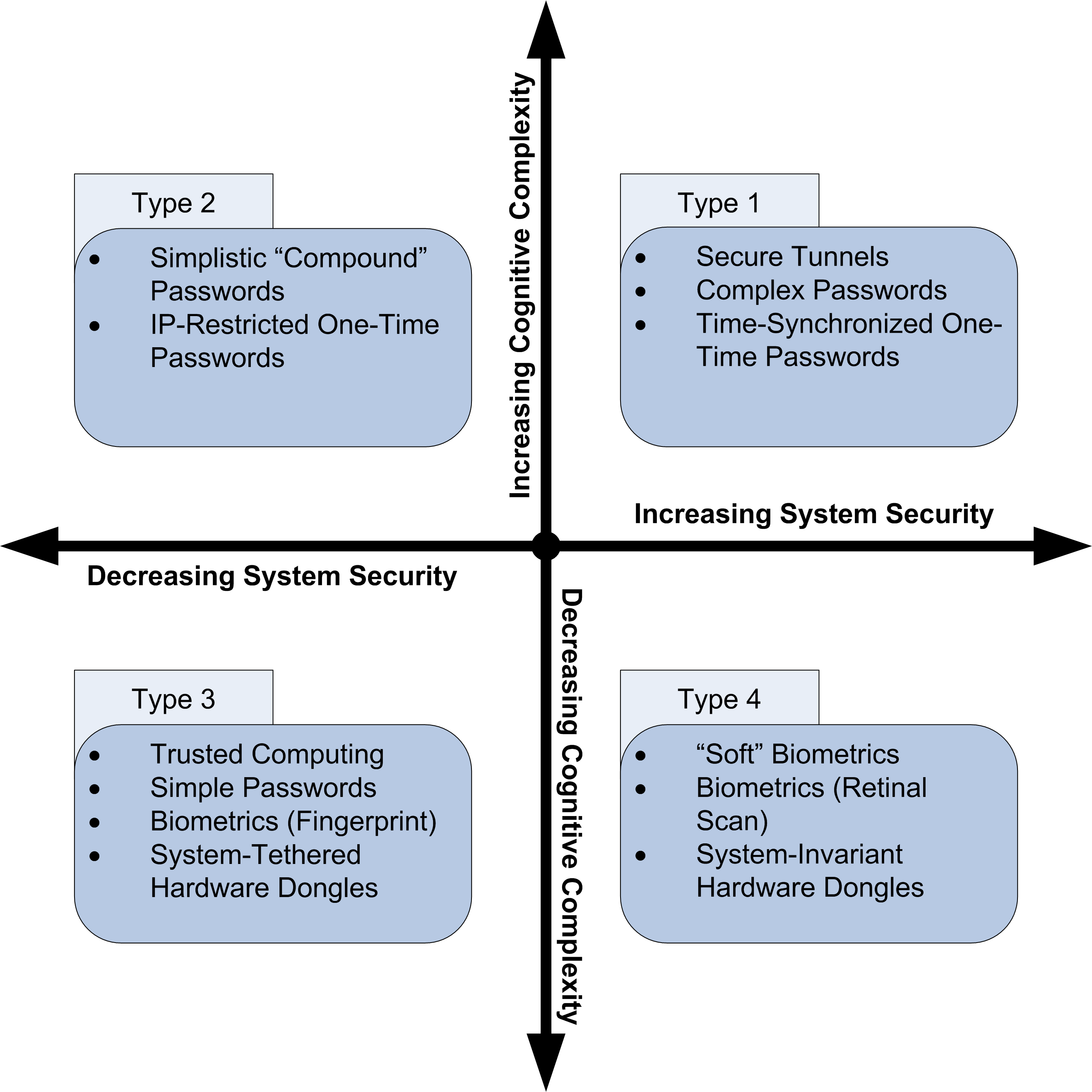}
    \label{fig:typology}
\end{figure}

Our typology, shown in figure~\ref{fig:typology} on page~\pageref{fig:typology}, describes authentication systems by their overall security and the level of cognitive complexity imposed upon a user. While it is simple to acknowledge that an ideal authentication system would possess a high measure of security while imposing a low degree of cognitive complexity on system users, few authentication mechanisms meet both criteria. Before addressing the ideal system, we will examine each of the other authentication system types.

\paragraph{Type 1: High Security, High Cognitive Complexity}
In situations requiring high levels of security, very sophisticated authentication systems have been deployed. Unfortunately, increased cognitive demand typically comes about as a consequence of this sophistication, creating what we call Type 1 authentication systems. Such systems are costly, not only because of the initial design and 
implementation costs, but also due to user training and lost productivity in the event that legitimate users are unable to access the system due to the cognitive complexity inherent in these systems. It is the user's inclination to reduce cognitive load via short-cuts (e.g., through written notes) that is of greatest concern, as it compromises the very asset security for which the system was designed.

Authentication mechanisms falling within the Type 1 designation include secure tunnels, highly complex passwords, and time-synchronized one time passwords (OTPs). The use of a secure tunnel--such as a Virtual Private Network (VPN)--creates cognitive demands through requiring an additional set of credentials for a user to remember. This additional set of credentials, in turn, increases the overall cognitive complexity of the authentication process. A second example of Type 1 authentication is highly complex passwords. These are usually discussed only in terms of resistance against brute-force attacks, but add their own problems through saddling users with highly irregular, impersonal, and wholly bizarre authentication strings. When users are presented with an extra set of credentials or a password such as \texttt{AfxZjY50!uzxQ43wRmH6}, it is natural that these users need some form of memory augmentation or assistance, resulting in potential security breaches.

Designed for remote access use, OTPs, such as S/Key\textsuperscript{TM} and OPIE (One-time Passwords In Everything); allow for one-time presentation of disposable credentials \cite{Lamport1981}. These credentials form a list, unique across users, and each password is valid for one and only one presentation, after which only the next password on the list permits authentication. Time-synchronized OTPs add further security through automatically expiring credentials after a specified duration of time. While this makes phishing more difficult, such OTPs place additional burdens on legitimate users. That is, unlike traditional passwords, they require that users not only have a record of passwords to present to the authentication mechanism, but that they are cognizant of their current location within the list.

In conclusion, Type 1 authentication mechanisms attempt to present the best possible asset security, but do so without regard for valid users. In other words, these mechanisms are tailored to preventing system access to an invalid user while paying little attention to the rejection of a valid user. In doing so, however, Type 1 mechanisms also provide an incentive for a valid asset user to reduce his/her cognitive load through scripting or storing hard-copies of passwords or password lists, thus compromising system security. These systems are inappropriate for the majority of IT installations and can easily overwhelm many organizations and system users.

\paragraph{Type 2: Low Security, High Cognitive Complexity}
A second type of authentication system offers little in the way of security to a determined attacker, while at the same time imposing cumbersome and unwieldy demands upon valid users. These Type 2 authentication routines represent the "worst-of-breed" in authentication mechanisms.

Examples of Type 2 authentication systems include simplistic "compound" passwords, Internet Protocol (IP)-restricted OTPs, and any security mechanism that puts more importance on point-of-entry obfuscation than security. The common trait shared by Type 2 systems is a cognitive imposition upon a valid user of an asset that far exceeds the security offered. 

Simplistic compound passwords are made up of two words joined together. It is quite possible for a user, when faced with an 8-character minimum length requirement, to present simple compound passwords (e.g. \texttt{forkorange} or \texttt{jim1947}). Such passwords can supply a measure of security, but unless the password is complex (Type 1), these passwords are fairly easy to uncover and can also be difficult for users to recall. 

While offering greater security than simplistic compound passwords, IP-restricted OTPs are not immune to phishing attacks given that potential hackers have time on their side. While time-sensitive OTPs (Type 1) mutate with the passage of time, IP-restricted one-time passwords do not; furthermore, IP addresses are easily spoofed. Consequently, an unused password allows invalid users time to launch a brute force attack against the password. Moreover, because credentials ultimately mutate with each presentation, users also tend to maintain their own "hard copy" of their password lists--and this paper list, consequently, may reveal precisely which password is active at a given time. 

In conclusion, Type 2 systems are the most insidious and inherently dangerous of the four types as they lull personnel into a false sense of security while simultaneously inconveniencing system users. In other words, they represent systems that evoke security through obscurity.

\paragraph{Type 3: Low Security, Low Cognitive Complexity}
A third type of authentication system provides for low overall security, but at a low cognitive cost to users. Although not recommended where security is vital, these Type 3 mechanisms are very user friendly. 

Examples within this category include trusted computing approaches, simple passwords, relatively weak biometrics such as the examination of a fingerprint, and system-tethered hardware dongles. In trusted computing there are no cognitive demands on the user as asset access is granted to anyone connecting to the asset from a particular physical or network location. Simple passwords, while conceivably user-specific, can be easily broken because users tend to pick passwords that are either trivial (e.g. \texttt{1234}) or otherwise non-novel (e.g., objects, names, etc.). Consequently, the level of protection afforded an asset guarded by a Type 3 authentication system is low. Simple biometrics, such as a fingerprint, provide greater security than simple passwords but are not fool-proof. For example, Matsumoto et al. demonstrated the inability of fingerprint authentication mechanisms to differentiate between "live" fingers and "false" fingers comprised of gelatin \cite{Matsumoto2002}. Finally, a "system-tethered" hardware dongle is a device that, by virtue of either its presence or contents, provides authentication. These dongles may be used in coordination with another authentication mechanism or alone. Examples include Bluetooth or RFID badges that provide authentication through their presence, or a hardware device that presents credentials stored on the device itself. Yet this solution is prone to circumvention as users may perpetually leave hardware dongles plugged in or otherwise accessible. 

In conclusion, Type 3 authentication mechanisms are very convenient from the user's perspective but provide little system security. Consequently, this category of authentication mechanisms should only be used where user access, rather than asset security, is the primary objective.

\paragraph{Type 4: High Security, Low Cognitive Complexity}
Finally, a fourth type of authentication system confers a high degree of system security while presenting a relatively low cognitive load to a user. These Type 4 systems are well suited for situations requiring high security.

Authentication mechanisms within this classification include "soft" biometrics, biometrics via retinal scanning, and system-invariant hardware dongles. "Soft" biometric mechanisms allow for a non-invasive examination of something that the user is (or does) as opposed to biometric mechanisms that leave behind latent information (e.g. fingerprints). Recently, keyboard input metrics, e.g., rhythm, tempo, etc., have been presented as examples of "soft" biometric mechanisms \cite{Chen2004, Guven2003}. Such "soft" biometric initiatives show promise and enjoy reduced cost through utilizing pre-existing hardware in the organization. Retinal scanners, in turn, are quick and intuitive, and benefit from properties of the retina. Biometric authentication via retinal scanning has traditionally been recognized as an exceedingly robust and secure access control mechanism, but also carries with it the advantage of a lack of physical, latent evidence of past users, unlike fingerprints. System-invariant hardware dongles are external devices that present a method to authenticate without directly storing the required credentials. The best example of a system-invariant hardware dongle comes in the form of challenge-response cryptographically-strong systems (e.g. RSA's SecureID system \cite{RSASecurity2006}), wherein the supplied information mutates and does not constitute the entire authentication mechanism. Along these lines, use of a 2-factor authentication token that does not carry any immediately "useful" information regarding the user, the token's use, or the system against which the token authenticates, tends to create an authentication scenario keeping with Type 4 systems, particularly for local log-in and some corporate networks. However, even these are not entirely immune to attack when used for Internet authentication \cite{Schneier2005}. Specifically, RSA's SecureID system generates a new authentication code every 60 seconds using a dongle-specific key. A user then uses this code along with a personal identification number (PIN) to authenticate. As the decision to authenticate or deny a user is made at the server, the token itself cannot be tampered with to send an "authenticate" signal to the system. Thus, the authentication system security is strong, while cognitive complexity to a user is low, as the PIN is reasonably short, and the longer code is provided through the token.

In conclusion, the key distinguishing feature of Type 4 authentication systems is the uniqueness and inherent complexity of their authentication credentials, either through personal identifiers (retinal scan) or through combining two different, independent mechanisms (dongle plus a PIN). This provides asset security without unduly imposing upon system users. 

\section{Conclusion}
The purpose of this paper was to develop and describe a typology of authentication systems. Accordingly, we classified examples of existing authentication mechanisms within the typology. We recognize that our typology does not take into account every authentication mechanism. Moreover, within-cell variations are likely. Not every authentication mechanism within a given system type will provide exactly the same levels of security and cognitive complexity, but these differences, both within- and between-cell, are testable. For example, one could test the cognitive complexity of an authentication system by users' ability to recall their password across sessions. Because what is cognitively complex may vary from person to person, such tests should include individual differences as potential moderating or mediating variables. One could also test the system's security via users' willingness to divulge their passwords to third-party users (unable to log in) and through one-way observation to see if users write their passwords down in order to recall them.

Our typology suggests that successful authentication systems must provide an appropriate balance between the need for security and the cognitive demands placed on valid users. If security is not an issue, then authentication should fall within Type 3 of the typology. Consideration of individual differences with respect to system authentication in high security environments leaves two options. One could strive to minimize the demands placed on users (i.e., Type 4 systems) or one could work to ensure that valid users have the cognitive capacity necessary to handle cognitively complex authentication mechanisms (i.e., Type 1 systems). Thus, the balance between security and users' abilities could be met either through technical solutions or through selection and training.

IT professionals recognize that the level of security in controlling system access must be balanced against the cost of such authentication systems. Sometimes the value of the asset one is trying to protect simply does not warrant elaborate authentication systems. We argue that the psychology of the user must also be considered. Too often technical requirements of these systems are pursued in the absence of user needs and requirements. Academically, there is already a movement to merge these two perspectives through the development of human computer interaction (HCI) programs; a movement we believe is needed. 

\bibliographystyle{ieeetr}
\bibliography{A-typology-of-authentication-systems}

\end{document}